\documentclass[10pt,conference]{IEEEtran}
\usepackage{amsmath,amsthm,amssymb,amsfonts}
\usepackage{algorithm}
\usepackage{algpseudocode}
\usepackage{graphicx}
\usepackage{caption}
\usepackage{subcaption}
\usepackage{hyperref}
\usepackage{cite}
\usepackage[font=small,labelfont=bf]{caption}
\usepackage[T1]{fontenc}
\usepackage{comment}
\usepackage{authblk}
\usepackage{algorithm}
\usepackage{algpseudocode}
\usepackage{booktabs}
\usepackage{tabu}

\usepackage[font={small,it}]{caption}

\usepackage[usenames,dvipsnames]{xcolor}
\usepackage{tcolorbox}
\usepackage{tabularx}
\usepackage{array}
\usepackage{colortbl}
\tcbuselibrary{skins}

\newcolumntype{Y}{>{\raggedleft\arraybackslash}X}

\tcbset{tab1/.style={fonttitle=\bfseries\large,fontupper=\normalsize\sffamily,
colback=yellow!10!white,colframe=red!75!black,colbacktitle=Salmon!40!white,
coltitle=black,center title,box align=base,freelance,frame code={
\foreach \n in {north east,north west,south east,south west}
{\path [fill=red!75!black] (interior.\n) circle (3mm); };},}}

\tcbset{tab2/.style={enhanced,fonttitle=\bfseries,fontupper=\normalsize\sffamily,
colback=white!10!white,colframe=black!50!black,colbacktitle=Salmon!40!white,
coltitle=black,center title, box align=base}}

\let\OLDthebibliography\thebibliography
\renewcommand\thebibliography[1] {
  \OLDthebibliography{#1}
  \setlength{\parskip}{1.8pt}
  \setlength{\itemsep}{1.8pt plus 0.3ex}
}

\let\oldref\ref
\renewcommand{\ref}[1]{(\oldref{#1})}

\IEEEoverridecommandlockouts

\hypersetup{
     citecolor = green
}

\begin{document}

\title{A refined SVD algorithm for collaborative filtering}
%surname "Duque Lopez" cannot be mixed.
\author{
  Marko Kabi\'c\\
  kabicm@student.ethz.ch\\
  ETH Z\"{u}rich
  \and
  Gabriel Duque L\'opez\\
  dugabrie@student.ethz.ch\\
  ETH Z\"{u}rich
  \and
  Daniel Keller\\
  daniel\textunderscore keller@student.ethz.ch\\
  ETH Z\"{u}rich
}

\pagestyle{plain}

\maketitle

\begin{abstract}
\noindent
Collaborative filtering tries to predict the ratings of a user over some items based on opinions of other users with similar taste. The ratings are usually given in the form of a sparse matrix, the goal being to find the missing entries (i.e. ratings). Various approaches to collaborative filtering exist, some of the most popular ones being the \emph{Singular Value Decomposition} (SVD) and \emph{K-means} clustering. One of the challenges in the SVD approach is finding a good initialization of the unknown ratings. A possible initialization is suggested by \cite{better-average-funk}. In this paper we explain how K-means approach can be used to achieve the further refinement of this initialization for SVD. We show that our technique outperforms both initialization techniques used separately.
\end{abstract}

\section{Introduction}
\noindent
\emph{A recommender system} (RS) tries to predict how a user would rate an item he  has not rated. A RS which bases its prediction on ratings of many users with similar taste is called \emph{collaborative filtering} (CF). This approach stems from the assumption that like-minded users tend to rate items in similar fashion.

Formally, we are given a sparse matrix $R = \left[r_{u,i}\right]_{M\times N}$ where each entry $r_{u,i} \in \left\{0, 1, 2, 3, 4, 5\right\}$ is the rating of a user $u$ for item $i$, with $1$ being the lowest and $5$ being the highest rate, whereas $0$ means that the user has not rated the item yet. Our task is to fill in the missing (i.e. zero-valued) entries in $R$.

CF can be seen as a matrix factorization problem \cite{cf-as-matrix-factorization}. Despite the numerous algorithms proposed, there is no unique algorithm that is superior in all possible scenarios \cite{comparative}. We focus on two classes of algorithms -- the ones based on Singular Value Decomposition (SVD) and the ones based on K-means clustering. We show how these two approaches can be combined to achieve better performances in order to outperform previously mentioned methods.

\subsection{Singular Value Decomposition method}
\noindent
This approach uses reduced SVD to approximate the ratings matrix $R$ with the product:
$$
R \approx U_{M \times C} \cdot D_{C \times C} \cdot V^T_{C \times N}
$$
Intuitively, $U$ depicts the affinity of users to each of the $C$ aspects of items, $D$ shows the presence of aspects in the data and $V$ shows the distribution of aspects inside of each item.  

The main challenge here is deciding how to initialize the missing entries in the  matrix $R$ before applying SVD. A naive solution would be to substitute each missing rating with the average rating of the item. However, in \cite{better-average-funk}, better initial approximation of the missing entries $r_{u, i}$ is proposed using the \emph{corrected averages}, which is described in detail in the section \ref{sec:method}. First we use the K-means algorithm to refine the corrected averages and get better initial approximations of the missing ratings, and then we apply SVD to get the final predictions of the missing entries in matrix $R$.

\subsection{K-means clustering method}
\noindent
K-means algorithm partitions the points into $K$ clusters. In CF, we usually treat either users (user-based filtering) or items (item-based filtering) as points, with ratings being their coordinates \cite{distance-paradigm-1,distance-paradigm-2}. This allows grouping points into clusters according to some distance or similarity metric. Despite that finding the optimal solution to this problem is NP-hard \cite{KmeanNPhard}, many practical heuristic versions have been proposed.

In this method, the problem of initializing the missing ratings still remains. Furthermore, the choice of the right metric might be crucial. In \cite{distance-paradigm-1,distance-paradigm-2}, different metrics are compared, with euclidean distance showing the best results. Additionally, this approach is highly dependent on the initial positions of centroids (centers of the clusters), because of the greedy nature of the algorithm which tends to get stuck in a local minimum. In particular, when applying K-means to CF, the normalization of ratings might be necessary, because of the non-uniformity of  users' ratings on different items \cite{normalization-importance}.

%The following paragraph needs some serious rewriting:
In \cite{pca-initialization}, it is shown that PCA dimension reduction might be particularly beneficial for K-means clustering, since the cluster subspace spanned by $K$ centroids is also spanned by the first $K-1$ principal components. Furthermore, the authors showed that in the cluster subspace, when compared to the original space, between cluster distances are \emph{almost} unchanged, while within  cluster distances might be reduced, which could make K-means algorithm more effective when run on the cluster subspace than on the original data. For this reason, we will use principal components as initial positions of centroids.

Once the clusters are found, there are different ways of using them to infer the final predictions of ratings. For example, if $r_{u, i}$ is missing, we can consider all users from the cluster to which user $u$ belongs, and take the most frequent rating among them (see \cite{distance-paradigm-1} for further details) to approximate $r_{u, i}$. Instead of this, we will take the weighted average of their ratings, with the weights being the inverses of the euclidean distance to $u$.

\section{Our method}\label{sec:method}
\noindent
We use the SVD approach to make the final predictions. As an initial filling of missing ratings, we use the weighted average of the following two approximations:
\begin{itemize}
\item corrected averages proposed by \cite{better-average-funk}. 
\item predictions from the K-means algorithm. 
\end{itemize}
\noindent
Later we show that this is indeed a refinement.

\subsection{Corrected averages}
\label{corrected_averages}
\noindent
Let $\overline{r}$ denote the average of all known ratings, $S_{u, \bullet}$ denote the set of all items that user $u$ has rated, $S_{\bullet, i}$ denote the set of all users who rated item $i$ and $S$ be the set of all pairs $(u,i)$ for which the value $r_{u, i}$ is known. Then missing entry $r_{u,i}$ can be approximated by:
\begin{equation}
r_{u, i} \approx \overline{r}_{\bullet, i} + \overline{\Delta r}_{u, \bullet},
\label{initialization}
\end{equation}
where $\overline{r}_{\bullet, i}$ is the corrected average of ratings of item $i$ and $\overline{\Delta r}_{u, \bullet}$ is the corrected average offset of user $u$ from $\overline{r}_{\bullet, i}$, i.e.
\begin{align*}
\overline{r}_{\bullet, i} = \frac{K_1 \cdot \overline{r} + \sum_{u \in S_{\bullet, i}}r_{u,i}}{K_1 + \left|S_{\bullet, i}\right|}; && \overline{\Delta r} = \frac{1}{\left|S\right|}\sum_{(u, i) \in S} \left(r_{u, i} - \overline{r}_{\bullet, i} \right)
\end{align*}
$$\overline{\Delta r}_{u, \bullet} = \frac{K_2  \cdot \overline{\Delta r} + \sum_{i \in S_{u, \bullet} }\left( r_{u, i} - \overline{r}_{\bullet, i} \right) }{K_2 + \left|S_{u, \bullet} \right|}$$
These equations are derived from the assumption that the real average tends to be closer to the global average than to the sparsely observed one. In other words, a prior is usually more reliable than an insufficient number of observations. In the extreme case, when user has not rated any of the items, his average offset will be equal to the global average offset $\Delta r$ of all users. Similarly for items not rated by any user. Therefore, this initialization mitigates the cold start problem: when nothing is known about a user or an item. Suggested values of the constants are $K_1 = 25$ and $K_2 = 10$. 

\begin{comment}However, in theory, these constants can be precisely defined as follows. Let $\mu$ be the vector of items' average ratings and \end{comment}

\begin{comment}
Let $R' = [r'_{u, i}]_{M, N}$ be a matrix with these approximations instead of the missing ratings, i.e.
\begin{equation}
{r'}_{u, i} = \begin{cases} 
r_{u,i}, \quad &r_{u, i} \neq 0 \\
\overline{r}_{\bullet, i} + \overline{\Delta r}_{u, \bullet}, \quad &r_{u, i} = 0
\end{cases}
\end{equation}
\end{comment}

\subsection{K-means predictions}
\label{k-means-predictions}
\noindent
Before running the K-means algorithm, we substitute each missing rating with the mean rating of the corresponding item and then normalize the matrix $R$, so that all the missing entries remain zero-valued after normalization as well. First, we use $K$ principal components as the initial centroids positions.

We treat users as points in an $N$ dimensional vector space, with ratings as coordinates. After we identified the clusters, we approximate each missing ratings $r_{u, i}$ as follows:
\begin{itemize}
	\item find the set $\mathcal{N}_u$ of the users belonging to the same cluster as user $u$.
	\item compute euclidean distances $d_{u, u'}$ between user $u$ and every neighboring user $u' \in \mathcal{N}_u$. 
	\item assign the weight $\omega_{u, u'}$ to each user $u'$ as follows\footnote{Adding $1$ to the distance serves as a protection from division by $0$.}:
$$\omega_{u, u'} = \frac{1}{1 + d^2_{u, u'}}$$
	\item approximate the missing rating as the weighted average of neighbors' ratings:
	\begin{equation}
		\label{weighted-ratings}
		r_{u, i} \approx \frac{\sum_{u' \in \mathcal{N}_u} \omega_{u, u'}\cdot r_{u', i}}{\sum_{u' \in \mathcal{N}_u} \omega_{u, u'}}
	\end{equation}
\end{itemize}

\subsection{Combining the approximations together}
\noindent
Observed corrected averages give better approximation than the usual mean, especially when a significant number of ratings are missing, alleviating the cold start problem, but might not be that sensitive to the similarities that might exist between users ratings. On the other hand, K-means algorithm is capable of detecting similarities, but yields unreliable results when many coordinates are unknown. Therefore, it might be reasonable to try to combine these two approximations together to get the refined initial approximations, before running SVD. One way to do that is to take the weighted average of these two approximations as an initial guess.

Let $r_{u, i}$ be the missing rating and $\lambda \in [0, 1]$ be the \emph{refinement factor}, then using \eqref{initialization} and \eqref{weighted-ratings}, we can approximate $r_{u, i}$ as follows:
$$
r_{u, i} = \lambda \cdot \left(\overline{r}_{\bullet, i} + \overline{\Delta r}_{u, \bullet}\right) + (1-\lambda) \cdot  \frac{\sum_{u' \in \mathcal{N}_u} \omega_{u, u'}\cdot r_{u', i}}{\sum_{u' \in \mathcal{N}_u} \omega_{u, u'}}
$$

Once we refined all the approximations of the missing ratings, we apply the SVD algorithm to refine our initial approximations even further.
\noindent
We provide the pseudocode of our algorithm:
\begin{algorithm}[H]
\caption{Refined SVD algorithm for CF}\label{refined-svd}
\begin{algorithmic}[1]
\Require $\text{ratings matrix: } R_{M, N}, \text{refinement factor: } \lambda \in [0, 1]$, \text{\# of clusters: } K and \text{approximation rank in SVD: } C
\Ensure $\text{predictions matrix } R^*_{M, N}$
\State $R^{corr}_{M, N} \gets$ approximate missing entries of $R$ using \eqref{initialization}
\State $R^{mean}_{M, N} \gets$ approximate missing entries of $R$ by the average ratings of corresponding items
\State $\sigma_{1, N} \gets$ standard deviation of each item's ratings in $R^{mean}$
\State $\mu_{1, N} \gets$ average ratings of each item's ratings in $R^{mean}$
\State $R^{mean} \gets \left(R^{mean} - \mu\right)/\sigma$ normalize $R^{mean}$
\State $R^{corr} \gets \left(R^{corr} - \mu\right)/\sigma$ normalize $R^{corr}$
\State $PC_{K, N} = PCA(R^{mean}, K)$ find first $K$ principal components
\State $Kmeans(R^{mean}, PC, K)$ find $K$ clusters in $R^{mean}$ starting from PC
\State $R^{\omega} \gets$ approximation of $R^{mean}$ given by \eqref{weighted-ratings}
\State $R^* \gets R^{mean}$
\For{all missing $(u, i)$ entries}
	\State $r^*_{u, i} \gets \lambda \cdot r^{corr}_{u, i} + (1 - \lambda) \cdot r^{\omega}_{u, i}$
\EndFor
\State $R^* \gets SVD(R^*, C)$ find $C-$approximation of $R^*$
\State $R^* \gets R^* \cdot \sigma + \mu$ denormalize the result back
\State \Return $R^*$
\end{algorithmic}
\end{algorithm}
\noindent
Input parameters $K$, $C$ and $\lambda$ need to be fine-tuned to achieve the best results, which is discussed in the following section.

\section{Results}\label{sec:results}
\begin{figure}[H]
\centering
\begin{subfigure}{.24\textwidth}
  \raggedleft
  \includegraphics[width=\textwidth]{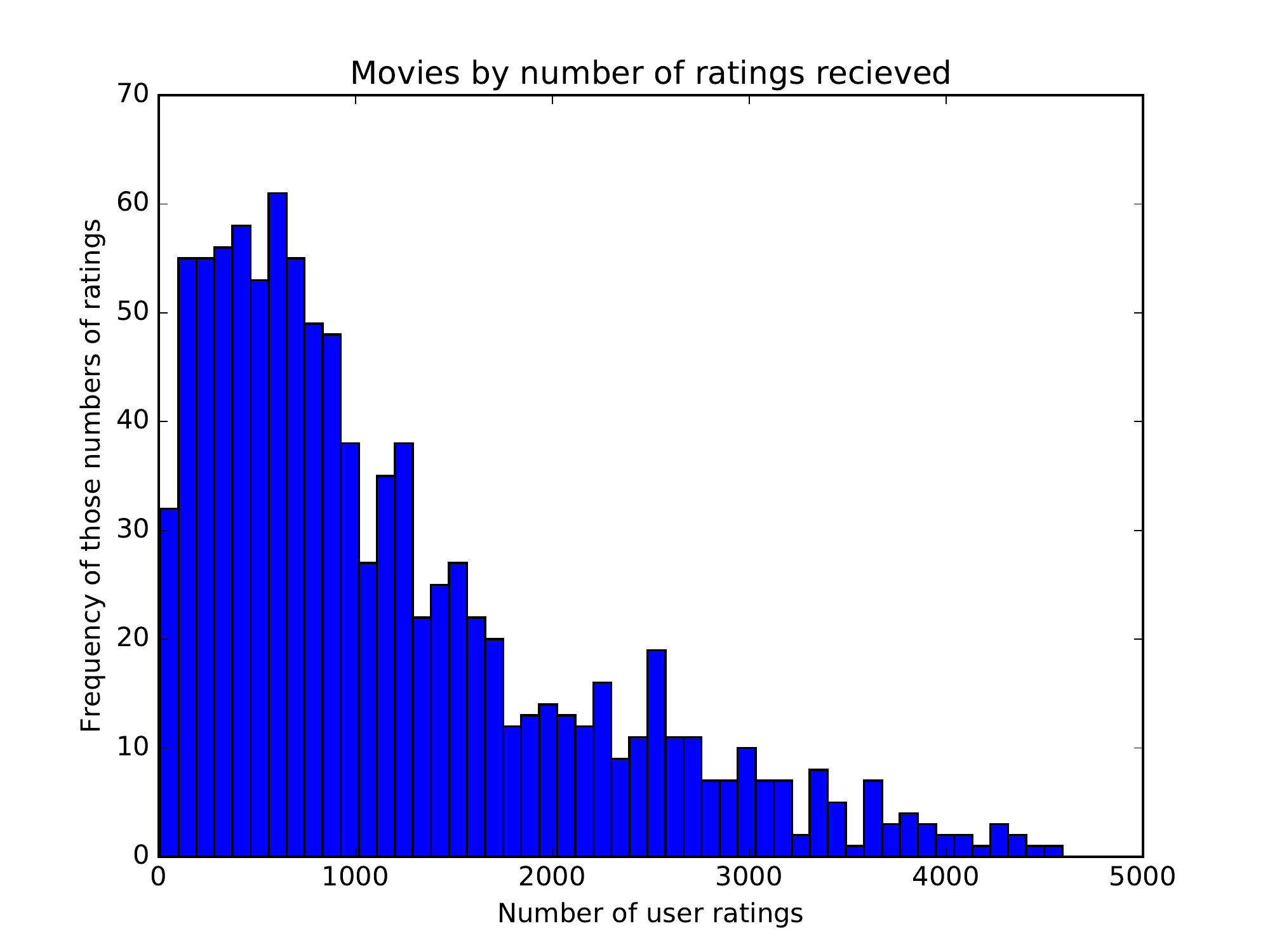}
  \caption{Number of ratings received by the movies}
  \label{fig:sub1}
\end{subfigure}%
\hfill
\begin{subfigure}{.24\textwidth}
  \raggedright
  \includegraphics[width=\textwidth]{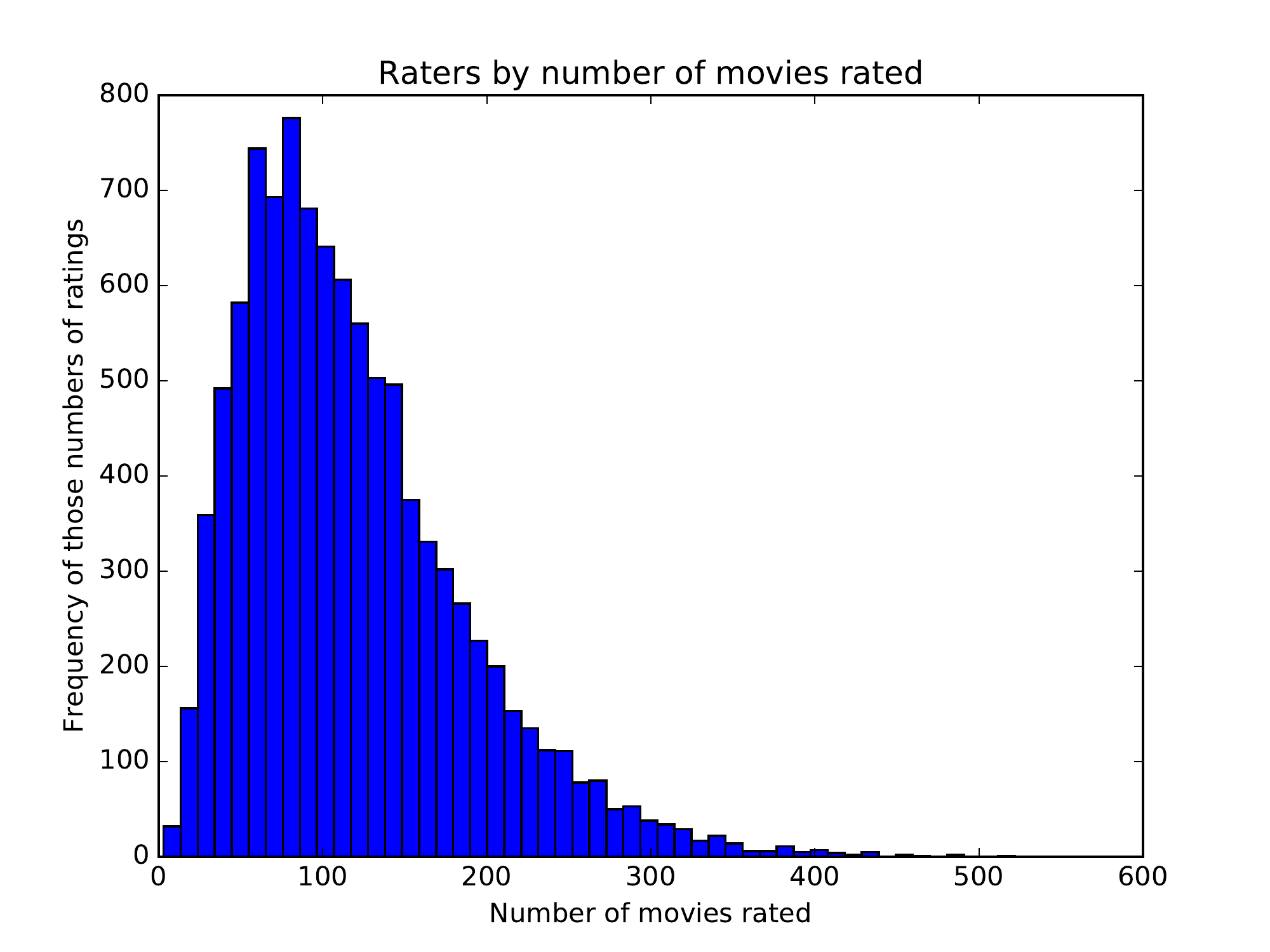}
  \caption{Number of ratings given by the users.}
  \label{fig:sub2}
\end{subfigure}
\caption{The available dataset is of excellent quality: the users or items which would be subject to the \emph{cold start} phenomenon have been filtered out. Both histograms are represented using 50 regular bins.}
\label{fig:dataHistogram}
\end{figure}

\noindent
In this section, we compare our refined SVD algorithm to similar CF methods under different settings, using RMSE as the quality metric. The data which was made available to us consisted of 1176952 ratings by 10000 users for 1000 items (a density of 0.117). This is a relatively dense matrix when compared to the reference data in the literature, such as the one used for the \emph{Netflix Grand Prize} (density: 0.0117) \cite{better-average} or the \emph{MovieLens} dataset used by \cite{distance-paradigm-1,distance-paradigm-2} (density: $\sim 0.0588$). The histograms in figure \ref{fig:dataHistogram} show the distributions  of the ratings.

The data set was split into the training set consisting of 80\% of known ratings, whereas the remaining 20\% were used to test the quality of predictions. The absence of almost non-rated items or inactive users made it impossible to test how our algorithm would cope with the problem of cold start, although the presence of corrected averages in our algorithm should make it flexible even in these cases. 

The constants in our algorithm are chosen such that the RMSE of our algorithm is minimized when applied to the test set, which yielded the following values:
$$\lambda = 0.6, \quad K = 7, \quad C = 28$$
We compared our algorithm to the baseline SVD (with naive initialization) and then to the SVD with only K-means initialization and SVD with only corrected averages initialization. We varied the number of clusters $K$ and approximation rank $C$ of SVD for each of the methods to achieve the lowest possible RMSE. Before discussing results separately, we first give the summary of results in the table \ref{tab:rawResults}.

\def\arraystretch{1.25}
\begin{table}[h!]
	\centering
	\begin{tabu}{|[1.5pt]l | c | c |[1.5pt]}
    	\tabucline[1.5pt]{-}
		\textbf{Method} & \textbf{Best RMSE} & \textbf{Approx.\ rank in SVD} \\
    	\tabucline[1.5pt]{-}
	    \itshape{Baseline SVD} & 1.01215 & 19 \\
        \hline
        \itshape{SVD with K-means $(K=7)$} & 1.00057 & 25 \\
        \hline
        \itshape{SVD with corrected averages} & 0.99439 & 24 \\
    	\tabucline[1.5pt]{-}
        \itshape{Refined SVD} $(\lambda = 0.6)$  & 0.99232 & 28 \\
    	\tabucline[1.5pt]{-}
	\end{tabu}
    \caption{The best results achieved for all mentioned algorithms.}
    \label{tab:rawResults}
\end{table}
%=======================================================
\subsection{Baseline SVD result}
\begin{figure}[H]
    \includegraphics[width=\linewidth]{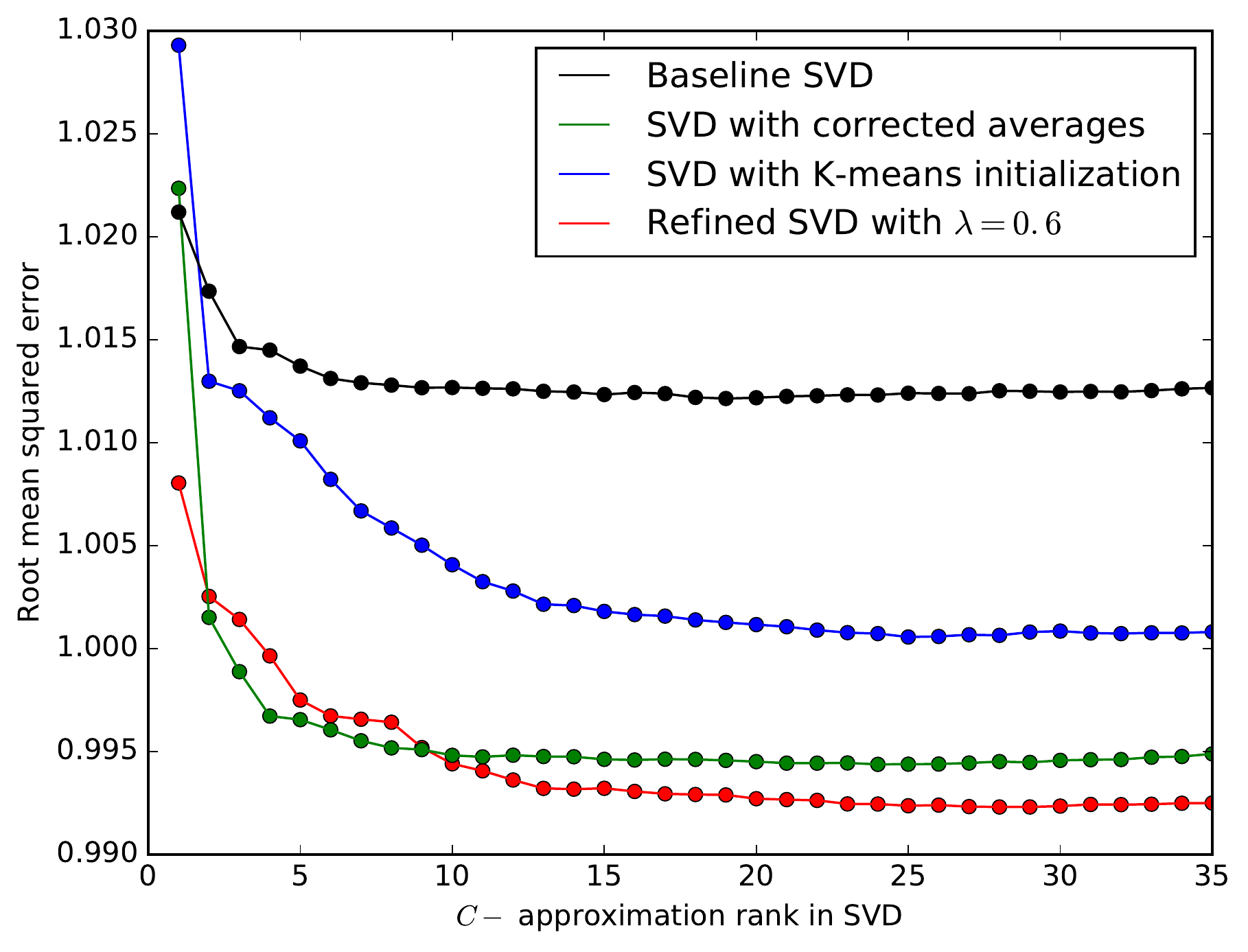}
    \caption{Comparison of the RMSE yielded by all mentioned methods in function of the number $C$ of singular values used for the reconstruction. }
    \label{fig:RMSE(k)_comp}
\end{figure}

\noindent
The first reference is certainly set by the SVD algorithm itself, naively initialized with the overall average rating as a placeholder for missing values. 
For this, we used the \texttt{numpy.linalg}, $LAPACK$-based implementation of the algorithm. In this setting, the best results are obtained for $C= 19$ singular values kept, as shown in the table \ref{tab:rawResults}, yielding a RMSE of $1.01215$. Because SVD is such a versatile algorithm, fast and well-optimized implementations are readily available. This ease of implementation and simplicity comes at the price of a comparatively high RMSE, which can also be seen in figure \ref{fig:RMSE(k)_comp}.

\subsection{SVD with corrected averages result}
\noindent
Using the corrected averages drastically improves the results (best RMSE: 0.99439), even when considering only a low-rank reconstruction. Furthermore, it only adds little computational overhead to the baseline method. This is indeed a powerful method which achieves small errors even when used alone. 

\subsection{SVD with K-means result}
\noindent
Despite a high computational cost and many degrees of freedom offering fine-tuning possibilities, recommendations produced by this hybrid collaborative filter never reach RMSE values smaller than 1. The computational complexity is dominated by computing the user-user distances in order to get the necessary weights that are used in this approach.

\subsection{Refined SVD result}
\begin{figure}[H]
	\centering
	\includegraphics[width=\linewidth]{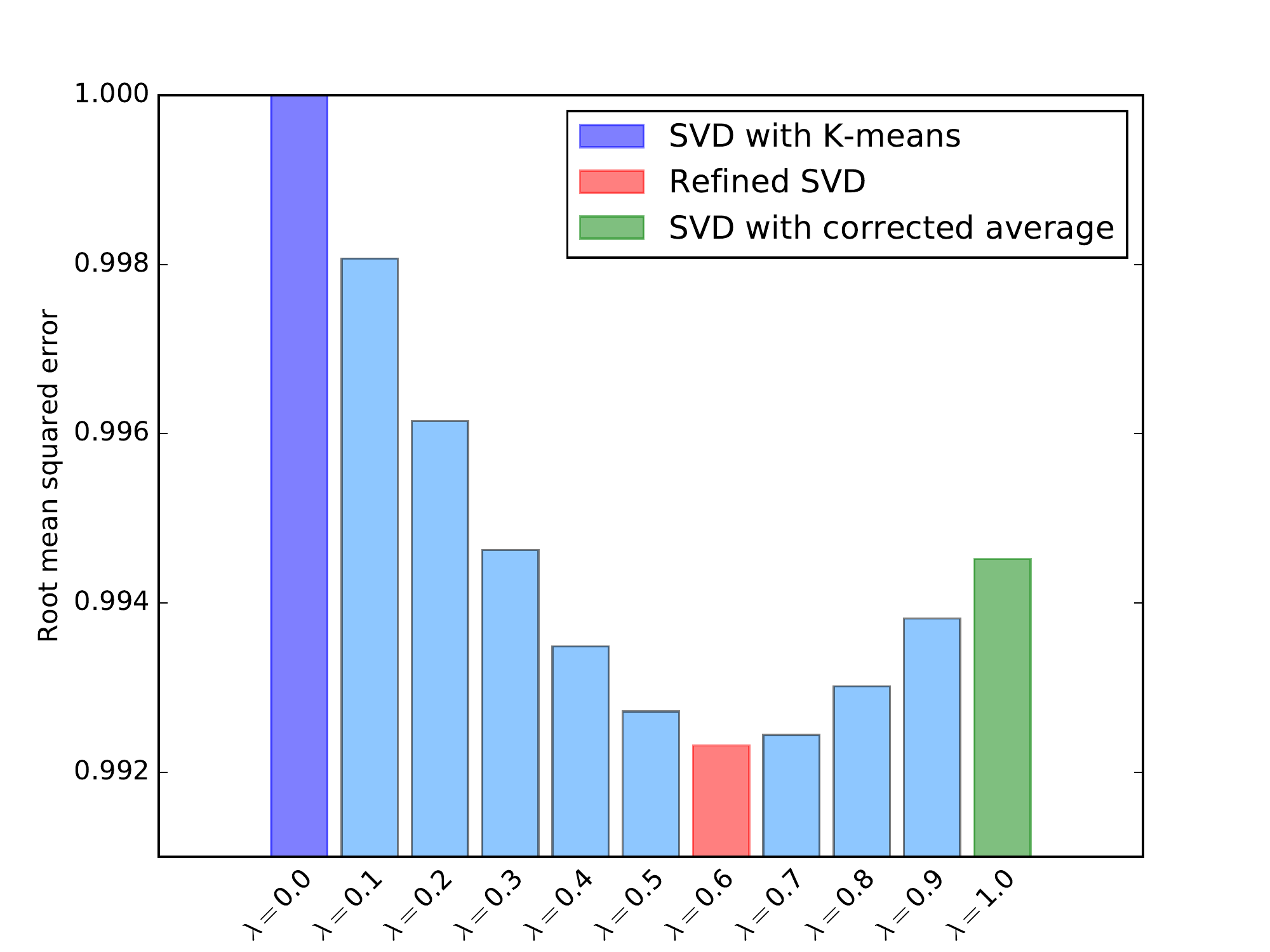}
    \caption{RMSE of the refined SVD algorithm plotted for different values of $\lambda$. The plot suggests that K-means initialization (blue) and corrected averages initialization (green) when combined together achieve better accuracy (red). }
	\label{fig:lambdaQuest}
\end{figure}

\noindent
Observe that for $\lambda = 0$ our algorithm initializes the missing entries only by using K-means algorithm, whereas for $\lambda = 1$ our algorithm initializes the missing entries only by using the corrected averages. The smallest errors that our algorithm produced for different values of $\lambda$ can be seen in figure \ref{fig:lambdaQuest}.
Notice that, when combined with SVD, initialization using only K-means algorithm (blue) and initialization using only corrected averages (green) both yield larger error when used separately than when combined (red), which indicates that our algorithm is indeed a refinement of both algorithms. 

However, comparing best possible errors alone cannot get the real insight into the performance of the algorithms since each algorithm achieves its optimum for a different value of the approximation rank $C$. Therefore, it is reasonable to ask how these algorithms perform when all of them are run using the same approximation rank $C$ in the SVD algorithm. In the following figure, we show how each of these algorithms behave for each of the approximation rank $C$.
We see that for small values of $C$, SVD with corrected averages even outperforms our algorithm, however, for higher values of $C$ it is not the case.

The linear combination of several initialization techniques obviously seems to improve the robustness of the predictions. One remarkable fact is the apparent convexity of the problem, with a single minimum located at $0.6$, and the surprising stability of this minimum, always close to $0.6$ even when the number of clusters and the approximation rank were changed, which indicates that, for large enough $C$ it might be independent of the number of clusters. Our novel collaborative filter outperforms all other considered methods in terms of accuracy. However, the matter of computational complexity still remains an issue.

\section{Discussion}\label{sec:discuss}

From table ~\ref{tab:rawResults}, we see that even though our algorithm is a refinement, the resulting RMSE is improved by only $\sim 0.2 \%$, with much additional computational cost, so one may ask whether it is really worth it. However, if we are concerned with computational power, there is one possible modifications to our algorithm that we considered.

Instead of the weighted average of neighboring ratings over distances, we can approximate each missing coordinate in K-means algorithm by the rating of the closest centroid, and then combine this with corrected averages. This approach saves much of a computational power and yields around $\sim 0.1 \%$ improvement. 

On the other hand, if we are more worried about the accuracy, we can hope for the better initial approximations by adding centroid rating as just one more term in our proposed approximation. However, not only does this add additional computational costs, but it also fails to achieve better results than our proposed algorithm.
This behavior is unsurprising, because the information contained in the centroids is very redundant to the one brought by the weighted average of ratings over distances in the same cluster. Therefore, if we want to further refine the initial approximation, it might make more sense to introduce some new method, different from K-means, that will bring new information as an additional term in our proposed approximation.

An interesting fact that we noticed while fine-tuning the constants is that when searching for different values of constants (number of clusters, approximation rank of SVD and $\lambda$) in order to find the global optimum, the constants seemed to be independant. In other words, if we fix any two of these constants and then optimize the RMSE, the remaining constant almost always had the same, or similar value in the optimum, like $\lambda$ always leading to good results when close to $0.6$, $C$ close to 7, etc. Similar stability was also noticed by \cite{better-average}.

\begin{comment}
Some even better results may be obtained if one tuned all constants simultaneously. In our approach, when a new constant had to be introduced, we ran the program for many values, keeping the best one. They aren't independent, so the introduction of a further one would mean having to adapt all previous ones in order to get the optimal result.  However, because of this being highly impractical, we almost never searched through sets of constants.
\end{comment}
\section{Conclusion}\label{sec:conclusion}

In this work, we show the importance of a good initialization values for the SVD decomposition through our novel approach to collaborative filtering. Blending some well-known techniques may greatly improve the final result. Adding a weight factor to the individual ratings within a cluster to even refine some inferred value results in a little better RMSE, at the price of a much more complex algorithm.

\nocite{*}
\bibliographystyle{IEEEtran}
\bibliography{refs}

\begin{comment}
CHANGELOG
===========================================================================
121: replaced K by C
131: TODO
176: added "THE" first, doubt about whether the sentence should be "First, we use..."
200: removed "ready". Is this an approximation or already the final prediction we are writing about?
\end{comment}

\end{document}